# The NuMI Neutrino Beam at Fermilab

Sacha E. Kopp

*Department of Physics, University of Texas*
*Austin, Texas 78712 U.S.A.*

**Abstract.** The Neutrinos at the Main Injector (NuMI) facility at Fermilab is due to begin operations in late 2004. NuMI will deliver an intense $\nu_\mu$ beam of variable energy 2-20 GeV directed into the Earth at 58 mrad for short (~1 km) and long (~700-900 km) baseline experiments. Several aspects of the design are reviewed, as are potential upgrade requirements to the facility in the event a Proton Driver is built at Fermilab to enhance the neutrino flux.

## INTRODUCTION

A new beam line facility is presently under construction at the Fermi National Accelerator Laboratory in Illinois. This facility, the "Neutrinos at the Main Injector" (NuMI) [1], will deliver an intense $\nu_\mu$ beam to what is planned to be a variety of experiments. The first experiment, MINOS [2], will attempt to confirm the atmospheric neutrino deficit seen by Super-Kamiokande and perform definitive spectrum measurements which demonstrate the effect of neutrino oscillations. The second, called MINERvA [3], is an experiment 1 km from the NuMI target to perform neutrino cross section measurements. A third proposal, called NOvA [4], would explore the phenomenon of CP violation in neutrinos.

NuMI is a tertiary beam resulting from the decays of pion and kaon secondaries produced in the NuMI target. Protons of 120 GeV are fast-extracted (spill duration 8.6 $\mu$sec) from the Main Injector (MI) accelerator and bent downward by 58 mrad toward Soudan, MN. The beam line is designed to accept $4\times10^{13}$ protons per pulse (ppp). The repetition rate is 0.53 Hz, giving ~$4\times10^{20}$ protons on target per year.

## MAIN INJECTOR BEAM

The MI is fed up to 6 batches from the 8 GeV Booster accelerator, of which 5 are extracted to NuMI. The Booster can deliver $5\times10^{12}$ *p/*batch, and efforts are underway to increase this number [5]. Studies of running multiple batches simultaneously in the MI have achieved accelerated beams of up to $3\times10^{13}$ protons. This number will likely improve after the shutdown because of the installation and commissioning of a new digital damper system [6] and a beam loading compensation system [7] for the MI.

Initiatives to produce higher intensity beam from the Main Injector without constructing a proton driver upgrade [8], including multiple batch 'stacking' of more than 6 Booster batches, could increase the beam intensity to NuMI by a factor of 1.5 at the expense of some emittance growth. A short run of synchronized transfer of 2 batches from Booster to Main Injector, followed by slip-stacking of the batches to a single batch of $7\times10^{12}$ protons, was accomplished in August [9]. However, only ½ the RF cavities had the beam loading compensation in place at that time, so the slip-stacking performance will likely improve in the coming year. The goal in 2005 is to implement slip-stacking for p-bars, and accelerate $3.3\times10^{13}$ protons per MI cycle ($8\times10^{12}$ ppp slip-stacked for antiproton production, $2.5\times10^{13}$ ppp for NuMI).

## PRIMARY BEAM TRANSPORT

The Main Injector beam is extracted by a set of three kicker magnets and three Lambertsons at the MI65 region of the MI. The transport line will maintain losses below $10^{-5}$ to reduce component activation, and below $10^{-6}$ loss in the region of the soil/rock interface to reduce activation of ground water. In anticipation of batch stacking in the MI, the momentum acceptance of the transport line will be $\Delta p/p = 0.0038$ at $40\pi$ mm-mr. Injection errors of ~1 mm lead to targeting errors of ~0.5 mm. The beam line has 24 BPM's, 19 dipole correctors, and 10 retractable segmented foil secondary emission detectors for measuring beam profile and halo [10].

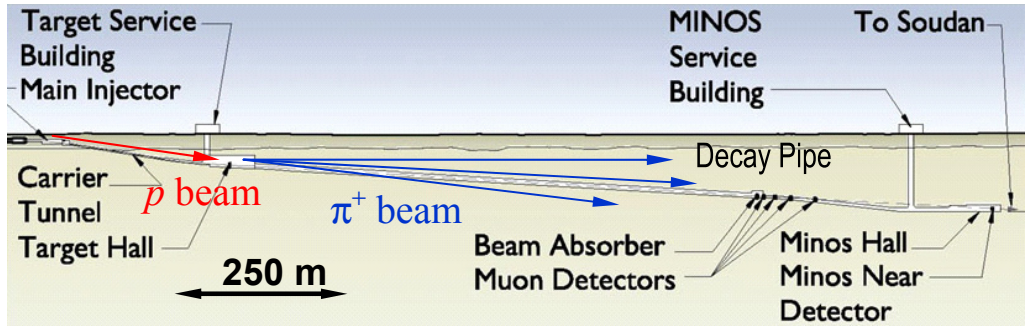

**FIGURE 1.** Elevation view of the NuMI facility, showing the 58 mrad downward inclination toward the Soudan Lab.

## TARGET AND HORNS

The primary beam is focused onto a graphite production target [11] of $6.4\times20\times940$ mm$^3$ segmented longitudinally into 47 fins. The beam size at the target is 1 mm. The target is water cooled via stainless steel lines at the top and bottom of each fin and is contained in an aluminum vacuum can with beryllium windows. It is electrically isolated so it can be read out as a Budal monitor [12]. The target has a safety factor of about 1.6 for the fatigue lifetime of $10^7$ pulses (1 NuMI year) given the calculated dynamic stress of $4\times10^{13}$ protons/pulse and 1 mm spot size. A prototype target was tested in the Main Injector in 1998 at peak energy densities exceeding that expected in NuMI. Studies indicate that the existing NuMI target could withstand up to a 1 MW proton beam if the beam spot size is increased from 1 mm to 2-3 mm [13].

The particles produced in the target are focused by two magnetic 'horns' [14]. The 200 kA peak current produces a maximum 16 kG toroidal field which sign- and momentum-selects the particles from the target. Field measurements on a horn prototype show the expected $1/r$ fall-off of the field to within a percent. The horns are designed to withstand $10^7$ pulses (1 NuMI year), and tests of the prototype horn have so far achieved this. The relative placement of the two horns and the target optimizes the momentum focus for pions, hence the peak neutrino beam energy. To fine-tune the beam energy, the target is mounted on a rail-drive system with 2.5 m of travel along the beam direction, permitting remote change of the beam energy without accessing the horns and target [15]. The neutrino spectra from several target position settings are shown in Figure 2.

Each of the two horns and the target are supported under shielding modules which are lowered by overhead crane down into the target cavern. Using the modules, failed horns or targets may be lowered remotely into a coffin for disposal. The horns, target, and support modules are cooled in the target cavern via recirculating air system flowing at 50 km/hr. Should significant proton intensity upgrades be pursued, this air system would require upgrade and the air flow volume better sealed so as to contain contamination from radioactive air molecules.

## DECAY VOLUME AND ABSORBER

The particles are focused forward by the horns into a 675 m long, 2 m diameter steel pipe evacuated to ~1 Torr. This length is approximately the decay length of a 10 GeV pion. The entrance window to the decay volume is a ellipsoidal bell-shaped steel window 1.8 cm thick, with a 1 mm thick aluminum window 1 m in diameter at its center where 95% of the entering pions traverse. The decay volume is surrounded by 2.5-3.5 m of concrete shielding to prevent contamination of groundwater. Twelve water cooling lines around the exterior of the decay pipe remove the 150 kW of beam heating deposited in the steel pipe and concrete.

Earlier plans to instrument the decay volume with a current-carrying wire (called the 'Hadron Hose' [16]) which would provide a toroidal field that continuously focuses pions along the decay pipe length, have been abandoned due to budget constraints. Such a device increases the neutrino flux by approximately 30%.

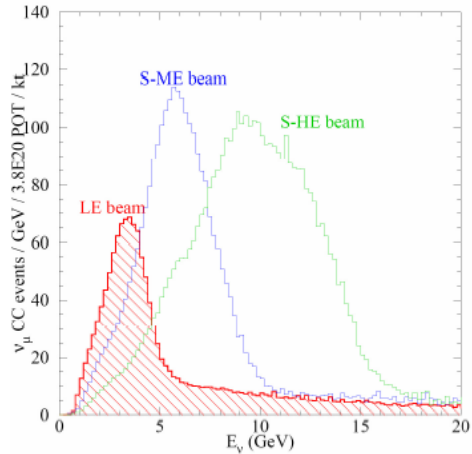

**FIGURE 2.** Neutrino beam energy spectra achieved at the MINOS experiment with the target in its nominal position inside the horn (LE), or retracted 1 m (ME) or 2.5 m (HE).

At the end of the decay volume is a beam absorber consisting of a 1.2×1.2×2.4 m$^3$ water-cooled aluminum core, a 1 m layer of steel blocks surrounding the core, followed by a 1.5 m layer of concrete blocks. The core absorbs 65 kW of beam power, but can sustain the full 400kW beam power for up to an hour in the event of mistargeting. In the event of a proton intensity upgrade the core would require no modification, but the steel blocks might require cooling.

## SECONDARY AND TERTIARY BEAM INSTRUMENTATION

Ionization chambers are used to monitor the secondary and tertiary particle beams [17]. An array is located immediately upstream of the absorber, as well as at three muon 'pits', one downstream of the absorber, one after 8 m of rock, and a third after an additional 12 m of rock. These chambers monitor the remnant hadrons at the end of the decay pipe, as well as the tertiary muons from $\pi$ and $K$ decays. When the beam is tuned to the medium energy configuration, the pointing accuracy of the muon stations can align the neutrino beam direction to approximately 50 µradians in one spill. In NuMI, the hadron (muon) monitor will be exposed to charged particle fluxes of $10^9$/cm$^2$/spill, ($10^7$/cm$^2$/spill). Beam tests of these chambers indicate an order of magnitude safety factor in particle flux over the rates expected in NuMI before space charge buildup affects their operation.

## OUTLOOK

Fermilab is in the midst of a 12-week shutdown whose purpose includes completion of the NuMI transport line in the MI tunnel. NuMI will commence operations with a short test run of the proton extraction line and instrumentation in December, 2004. The target hall installation will be complete Jan. 1, 2005, whereupon commissioning of the neutrino beam with target, horns, and tertiary beam instrumentation will commence. In the coming year it is hoped that the Booster/MI complex will be commissioned to deliver $3.3\times10^{13}$ protons/cycle ($2.5\times10^{13}$ ppp for NuMI). Significant experience will also be gained regarding the adequacy of component cooling and of the shielding for future proton intensity upgrades, such as the proton driver.

## ACKNOWLEDGEMENTS

It is a pleasure to acknowledge the many who have contributed to the NuMI facility from the Fermilab Accelerator, Technical, and Particle Physics Divisions, as well as from the MINOS Collaboration and IHEP-Protvino.